\documentclass{article}
\usepackage{graphicx}
\usepackage{multirow}
\usepackage{url}

\usepackage{tabularx}
\usepackage{enumerate}
\usepackage{hyperref}

\newcommand{\ignore}[1]{}
\newcommand{\revised}[1]{}
\newcommand\comment[1]{}

\begin{document}

\title{\Large \bf Unauthorized Cross-App Resource Access on MAC OS~X and iOS}


\author{
{\rm Luyi Xing{*}, Xiaolong Bai, Tongxin Li, XiaoFeng Wang}\\
Kai Chen, Xiaojing Liao, Shi-Min Hu, Xinhui Han\\
{*}luyixing@indiana.edu, Indiana University Bloomington
} 
       
\date{}

\maketitle

\subsection*{Abstract}

On modern operating systems, applications under the same user are separated from each other, for the purpose of protecting them against malware and compromised programs. Given the complexity of today's OSes, less clear is whether such isolation is effective against different kind of cross-app resource access attacks (called \textit{XARA} in our research).  To better understand the problem, on the less-studied Apple platforms, we conducted a systematic security analysis on MAC OS~X and iOS.  Our research leads to the discovery of a series of high-impact security weaknesses, which enable a sandboxed malicious app, approved by the Apple Stores, to gain unauthorized access to other apps' sensitive data. More specifically, we found that the inter-app interaction services, including the keychain, WebSocket and NSConnection on OS~X and URL Scheme on OS X and iOS, can all be exploited by the malware to steal such confidential information as the passwords for iCloud, email and bank, and the secret token of Evernote.  Further, the design of the App sandbox on OS~X was found to be vulnerable, exposing an app's private directory to the sandboxed malware that hijacks its Apple Bundle ID.  As a result, sensitive user data, like the notes and user contacts under Evernote and photos under WeChat, have all been disclosed.  Fundamentally, these problems are caused by the lack of app-to-app and app-to-OS authentications.  To better understand their impacts, we developed a scanner that automatically analyzes the binaries of OS~X and iOS apps to determine whether proper protection is missing in their code.  Running it on hundreds of binaries, we confirmed the pervasiveness of the weaknesses among high-impact Apple apps. Since the issues may not be easily fixed, we built a simple program that detects exploit attempts on OS~X, helping protect vulnerable apps before the problems can be fully addressed. We further discuss the insights from this study and the lessons learnt for building a securer system. 

\section{Introduction}
\label{sec:introduction}
The pervasiveness of computing technologies and emerging security threats they are facing have profoundly changed the security designs of modern operating systems (OS).  Moving away from the traditional threat model in which all applications (\textit{app} for short) under the same user trust each other with their information assets, today's OSes tend to separate those apps and their resources, in an attempt to prevent a malicious or compromised program from causing damage to others.  This has been achieved through a variety of \textit{app isolation} mechanisms:  each app is confined in its partition with a minimum set of privileges, called \textit{sandbox}, and needs to explicitly require additional capabilities (e.g., access to camera, audio, etc.) from the OS or the user. Such a security model has been adopted by most mainstream systems, including Windows, MAC OS~X, Android, iOS, etc. With its popularity, the effectiveness of the technique, however, has still not been fully understood, due to the complexity of a modern OS, which makes comprehensive protection challenging.

\vspace {3pt}\noindent\textbf{Unauthorized cross-app resource access}.  Recent studies show that sandboxed Android apps can still get access to other apps' resources and acquire system capabilities without proper authorization~\cite{Xing:2014:UYA:2650286.2650760}. For example, the developer could accidentally make public an app's interface for interprocess communication (IPC), through which its internal \textit{service} or \textit{activity} can be triggered by a message (called \textit{Intent}) from an unauthorized app to acquire sensitive data~\cite{wang2013unauthorized}\ignore{domain-crossing paper} or elevated privileges (e.g., access to audio, GPS, etc.)~\cite{Davi:2010:PEA:1949317.1949356, Lu_chex:statically, felt2011permission}. Fundamentally, the problem is caused by the migration of the threat model and the transitional pain that it comes with: both the OS designer and the app developer are less used to the mindset that all apps, even when they all belong to the same user, should treat each other as untrusted, and proper security checks should always be performed in all aspects of app-to-app and app-to-system interactions.

In those attacks, malicious code under some isolation constraints manages to gain access to other apps' resources or affect the way they are used by legitimate apps, when it is not authorized to do so. We call such a security threat \textit{unauthorized cross-app resource access} or \textit{XARA}. Although specific instances of XARA are found on the Android platform, less known is whether it is indeed a generic issue. Particularly, we do not know whether app isolation works effectively on MAC OS~X and iOS, which are widely considered to be securer than Android.  These operating systems offer unique mechanisms to confine apps and support cross-app interactions, very different from those provided by Android.  Specifically, the construction of Apple sandboxes is significantly different from that of Android, in which each app is given a unique User ID (UID), allowing the Linux user protection to separate the apps.  In contrast, an Apple app is identified by its Apple ID, which contains a \textit{Bundle ID} (BID) token used by the OS to enforce sandbox policies. The uniqueness of the token is ensured by the Apple Store. Also, OS~X supports complicated cross-app resource sharing. For example, its keychain service allows multiple apps to share credentials among them through an access-control list (Section~\ref{subsec:keychain}), which is not supported on other systems like Android. In addition to cross-app resource sharing, other cross-app interactions, i.e., IPC on Apple platforms, also differ from those on Android. Examples include \textit{NSConnection} that shares objects between apps on OS~X and the \textit{URL Scheme} uniquely associated with one single app, for launching it with an URL\footnote{\small On Android, an Intent-based Scheme is different as it can be connected to multiple apps, which the user can choose once the scheme is triggered.}. So far, little has been done to understand whether the construction of app isolation on Apple platforms is secure and whether its cross-app mechanisms can bring in XARA risks never known before.

\vspace {3pt}\noindent\textbf{Our work}. We conducted the first study on the XARA risks of Apple's isolation mechanisms, and discovered surprising security-critical vulnerabilities: major cross-app resource-sharing mechanisms (such as keychain) and communication channels (including WebSocket, NSConnection and Scheme) turn out to be insufficiently protected by both the OS and the apps using them, allowing a malicious program to steal from these apps sensitive user data; also the BID-based sandbox construction is found to be less reliable than expected, and its resource-sharing mechanism can be exploited by the malicious app to break the sandbox confinement on OS~X, gaining full access to other apps' directories (called \textit{container}). Note that not only does our attack code circumvent the OS-level protection but it can also get through the restrictive app vetting process of the Apple Stores, completely defeating its multi-layer defense.

Looking into the root cause of those security flaws, we found that in the most cases, neither the OS nor the vulnerable app properly authenticates the party it interacts with\ignore{talks to}. To understand the scope and magnitude of this new XARA threat, we developed an analyzer for automatically inspecting Apple apps' binaries to determine their susceptibility to the XARA threat, that is, whether they perform security checks when using vulnerable resource-sharing mechanisms and IPC channels, a necessary step that has never been made clear by Apple. In our study, we ran the analyzer on 1,612 most popular MAC apps and 200 iOS apps, and found that more than 88.6\% of the apps using those mechanisms and channels are completely exposed to the XARA attacks (Section~\ref{subsec:impact}), and every app's container directory has been fully disclosed. The consequences are dire: for example, on the latest Mac OS~X 10.10.3, our sandboxed app successfully retrieved from the system's keychain the passwords and secret tokens of iCloud, email and all kinds of social networks stored there by the system app Internet Accounts, and bank and Gmail passwords from Google Chrome; from various IPC channels, we intercepted user passwords maintained by the popular 1Password app (ranked 3rd by the MAC App Store) and the secret token of Evernote (ranked 3rd in the free ``Productivity'' apps); also, through exploiting the BID vulnerability, our app collected all the private notes under Evernote and all the photos under WeChat\ignore{ even replaced the extension installer of Dashlane, another popular password manager, with malicious code}.  We reported our findings to Apple and other software vendors, who all acknowledged their importance.  \ignore{The video demos of our attacks and our communication with the related parties are posted on a private website~\cite{supporting}\footnote{\small We do not track the visitor.}}. We also built an app that captures the attempts to exploit the weaknesses.

Our study also shows that this XARA hazard is indeed general, across different platforms. Even though iOS drops many useful functionalities of OS~X (e.g., keychain's access control list for sharing passwords or tokens across apps) and therefore less vulnerable, it is still not immune to the threat.  Particularly, its major IPC channel, Scheme, is equally subject to the hijacking attack we discovered on MAC OS~X (Section~\ref{subsec:scheme}). Further, the WebSocket problem (Section~\ref{subsec:IPC}) actually comes from HTML5, which happens when a browser extension is connecting to a local program.  We found that the same attack can also succeed on iOS and Windows. Interestingly, compared with OS~X and iOS, Android looks pretty decent in terms of its protection against the XARA threat: at the very least, it offers a mechanism to protect its Intent-based IPC, through assigning a \texttt{private} attribute to the service and activity or guarding them with permissions, which are missing on the Apple platforms.\ignore{ On the other hand, subtle XARA problems are still there, even when the app developer makes proper use of the OS protection: for example, we found that a malicious app can use the authority name of its content provider to block the installation of antivirus app \textit{360 Security - Antivirus FREE} (Section~\ref{sec:discuss}).} We further discuss the lessons learnt from our study, particularly the need for clarifying the responsibilities for protecting a cross-app mechanism between the OS provider and the app developer, and present key principles for avoiding XARA pitfalls when building new systems (Section~\ref{sec:discuss}).

\vspace {3pt}\noindent\textbf{Contributions}.  The contributions of the paper are outlined as follows:

\vspace {2pt}\noindent$\bullet$\textit{ New understanding of the XARA threat}.  We are the first to identify the generality of the XARA problem and systematically investigate the threat on the Apple platforms.  Our study brings to light a series of unexpected, security-critical flaws that can be exploited to circumvent Apple's isolation protection and its App Store's security vetting.  The consequences of such attacks are devastating, leading to complete disclosure of the most sensitive user information (e.g., passwords) to a malicious app even when it is sandboxed. Such findings, which we believe are just a tip of the iceberg, will certainly inspire the follow-up research on other XARA hazards across platforms. Most importantly, the new understanding about the fundamental cause of the problem (Section~\ref{sec:discuss}) is invaluable to the development of better app-isolation protection for future OSes.


\vspace {2pt}\noindent$\bullet$\textit{ New effort to mitigate the threat}. We developed new techniques for identifying the apps vulnerable to the XARA threat, and the attempts to exploit them during an operating system's runtime. \ignore{We further discuss potential OS-level protection against this new type of security threat.}

\vspace {3pt}\noindent\textbf{Roadmap}.  The rest of the paper is organized as follows: Section~\ref{sec:background} provides the background information for our research and the assumptions we made; Section~\ref{sec:attack} elaborates the security analysis we performed on OS~X and iOS, and the security problems we discovered; Section~\ref{sec:detectandmeasure} describes the design and implementation of the automatic analyzer, the findings made by running the tool on popular apps and the app-level mitigation we developed; Section~\ref{sec:discuss} highlights the lessons learnt from our study; Section~\ref{sec:relatedwork} reviews the related prior research and Section~\ref{sec:conclude} concludes the paper.

\vspace {-8pt}
\section{Background}
\label{sec:background}

\ignore{As discussed earlier, modern OSes utilize various isolation techniques to separate apps from each other and limit the damage they may inflict on the system and other apps. }In this section, we describe how app isolation techniques work on popular systems like Android, MAC OS~X and iOS, the way they handle inter-app communication and security risks that come with such a strategy. Also, we present the adversary model underlying our study.

\vspace {3pt}\noindent\textbf{App sandboxing}. App sandboxing plays a critical role in the Android security architecture. Each Android app is given a unique UID and runs as the user. Sensitive resources are assigned to Linux groups such as GPS, Audio, etc. This treatment automatically isolates one app from others under the Linux user and process protection. To access system resources, an app needs to request permissions from the OS or the user.  A permission can also be defined by the app for sharing its resources with authorized parties (those with the permission) through the interfaces like content providers, Intent receiver, etc.

The Apple sandbox first appears on MAC OS~X, which utilizes the TrustedBSD mandatory access control framework to enforce its security policies at the system-call level. Since OS~X 10.7.5 Lion, all apps submitted to the MAC App Store are required to be sandboxed, with some exceptions given to those that need to run as native code. On the OS side, a service called \textit{Gatekeeper} blocks the apps not signed by either the Apple Store or a trusted developer from being installed\footnote{\small This setting can be turned off.}.  This ensures that with proper security configurations, most apps running on a MAC device are under the sandbox confinement. In the meantime, OS~X maintains its compatibility with the traditional OS security design, hosting trusted native programs that run with the user's privileges.\ignore{ The approach is necessary for accommodating the desktop applications with complicated functionalities.} On iOS, however, apps are much simpler (e.g., without intensive document operations) and can therefore all be sandboxed.

Unlike Android, which isolates an app solely based upon its UID, the Apple platforms just utilize UIDs to classify apps into groups. For example, on OS~X all the apps from the MAC app store operate under\ignore{are given} the UID of the current OS user, and those on iOS \ignore{all operate} under the user \texttt{mobile}. On these platforms, separation is actually enforced through the TrustBSD's API interpositions. Each app is identified by its \textit{Apple ID}, a two-part string that consists of a \textit{Team ID} Apple assigns to the app developer, and a Bundle ID supplied by the developer: for example, \texttt{A1B2C3D4E5.com.apple.mail} where the first part is the Team ID and the rest components form the BID. Any app submitted to the Apple Stores goes through a verification process that among other things, ensures the uniqueness of the app's BID. On OS~X, this identity also serves as the name of the app's \textit{container} directory. Every sandboxed app on the Apple platforms is given a container when it is first launched. The directory is used to hold the app's internal data and cannot be accessed by other sandboxed apps from different developers.

An app within the sandbox has only limited privileges. By default, it can only read and write files within its container and some public directories. This policy is enforced by checking the developer's signature on the app against an access-control list (ACL) associated with each directory (see Section~\ref{subsec:bundleid}).  Also, it is not allowed to access network sockets, built-in camera, microphone, printer and other resources. Whenever use of such resources becomes necessary, the app explicitly requires them by declaring a set of \textit{entitlements} within its property file (called \textit{plist} file, very much like the Android manifest file). Each entitlement is a key-value pair that identifies a specific capability (e.g., access to camera). They are reviewed by the Apple Stores to determine whether the capabilities should be granted. For some capabilities, such as access to GPS locations, camera, etc., the OS further asks for the user's permission during the app's runtime.


\vspace {3pt}\noindent\textbf{IPC on the Apple platforms}. Among the small set of operations that a sandboxed app is allowed to do by default is the capabilities to perform some types of interprocess communication. OS~X supports a variety of IPC channels, including traditional UNIX ones (e.g., pipe, UNIX domain socket, shared memory) and Apple-specific mechanisms like distributed objects, NSConnection in particular, and URL schemes.  More specifically, a sandboxed app, without any additional permission, can create an \texttt{NSConnection} server object, vend it and register with the OS the name of the object. This allows another app (i.e., an NSConnection \textit{client}) to communicate with the server after obtaining from the OS a proxy for the server object using its name. Specifically, through the proxy, the NSConnection client gets the vended object from the server. The NSConnection mechanism allows the client to invoke methods of the vended object and access its variables as if the object existed in the client process.  To this end, the client app needs to declare an entitlement \textit{com.apple.security.temporary-exception.mach-lookup.gl\\obal-name} in its plist.


Socket-based IPC is also available on OS~X. To use it, sandboxed apps need to claim the network capability in their plists. Another unique IPC mechanism for both OS~X and iOS is \textit{Scheme}: an app can invoke another specific app to work on a task with a URL click if the latter registers with the OS the scheme part of the URL. For example, the URL \url{yelp://search?terms=Coffee}, once triggered, let one app launch the Yelp app to search for ``Coffee'' nearby. Here, the ``\url{yelp://}'' part is a scheme. Although this mechanism is also used on Android, which has been implemented using Intent, it is different from that for OS~X and iOS since Apple's OSes only allow one single app to be associated with a scheme on a device, while on Android, the user is asked to choose a scheme's owner when there are more than one.  This major difference enables our scheme hijacking attack (Section ~\ref{subsec:scheme}) which, however, does not pose a threat on Android. \ignore{it is the major IPC for iOS apps.} To register a scheme, an Apple app needs to register it with the OS.  This is done on OS~X and iOS by simply declaring the scheme in the app's plist file. Such a channel can be used by any sandboxed app without specifying any entitlement.

\vspace {3pt}\noindent\textbf{Adversary model}. In our research, we studied what an isolated app can still do to collect sensitive data and utilize critical sources that belong to other apps, when it is not entitled to do so. For this purpose, we assume that malicious apps are submitted to the Apple Stores, which puts them to the test of Apple's restrictive review process.  In the case that they get published, the apps are supposed to be installed by the user who also runs security-critical apps on her device (laptop or smartphone). This is realistic, since apps downloaded from the Apple Stores are widely considered to be trusted, and particularly, almost all of them are confined within the sandboxes.  For the malware installed in this way, we assume that they are isolated and only granted a small, inconspicuous set of capabilities: in addition to what are offered by the OSes by default, they may need the networking permission (only for the attack in Section~\ref{subsec:IPC}) or that for the IPC client (for the NSConnection attack). Note that these entitlements are among the most innocent ones\ignore{, given the fact that most apps nowadays require network connections and IPC has been extensively used}.

\vspace {-7pt}
\section{XARA Menaces}
\label{sec:attack}

In our research, we conducted a systematic study on the XARA threat over the Apple platforms, MAC OS~X in particular.\ignore{ Our purpose is to understand whether the current app isolation mechanisms can prevent unauthorized programs from accessing other apps' sensitive resources.  To this end,} Our focus is on how inter-app interaction channels and services are protected under the sandboxing model, and how isolation has been enforced on untrusted apps.\ignore{  Also we notice that oftentimes, the onus of resource protection can be on either the OS or the app: for example, in the case that some IPC channels turn out to be unguarded by the app sandbox, the developer could make up for this weakness by building into her app additional authentication and authorization, even though this has never been suggested by Apple.  Therefore, understanding of the significance of OS-level lapses also requires identifying and analyzing the apps affected by the problem\ignore{, to ensure that it is indeed vulnerable. This has been done in our research through automatic analysis of Apple apps' binary code (Section~\ref{sec:detectandmeasure})}.}  Following we elaborate our findings, including the security-critical flaws we discovered in the OS~X keychain, BID-based separation as well as various IPC channels, i.e., NSConnection, WebSocket and Scheme \ignore{weaknesses} on both MAC OS~X and iOS. Note that \textit{all} our attack apps were uploaded to the Apple App Stores and passed their inspections\footnote{\small To avoid causing damages to Apple users, \ignore{our apps do not include the attack payloads that send sensitive data out of a device. Also, }after the apps were confirmed to be approved, we immediately removed them from the Apple Store.}.

\subsection{Password Stealing}
\label{subsec:keychain}

On the Apple platforms, a sandboxed app by default is still allowed to access some security-critical services.  A prominent example is Apple's \textit{keychain}. Keychain is Apple's credential management service, through which an app can store the user's passwords, secret keys and certificates there.  These credentials will then be automatically used by authorized apps after the user ``unlocks'' the keychain through entering her password, in a way similar to the transparent single-sign-on authentication (though more powerful) from the user's point of view. When the keychain is locked, all the credentials there are encrypted and no one can access their content. The keychain service running on OS~X is powerful, supporting multiple keychains, explicit and implicit unlocking and complicated access control. Particularly, a default keychain is created for each user account and serves most system services and many popular apps.  It is automatically unlocked whenever the user logs in, if its password is identical to that for login.

Although keychain is not part of the Apple sandbox, it can be viewed as a secure storage system that provides a strong isolation between apps. Even when it is unlocked, each app cannot touch another's keychain item unless this is permitted by the item's creator, as specified by its ACL. For a sandboxed app, other apps' items are very much like being inside their individual container directories, which it is not allowed to access.  However, we show how a subtle design weakness enables the malicious code to bypass the isolation and steal user credentials from other apps.

\begin{figure}[h]
\centering
\vspace{-8pt}
\includegraphics[width=0.8\linewidth]{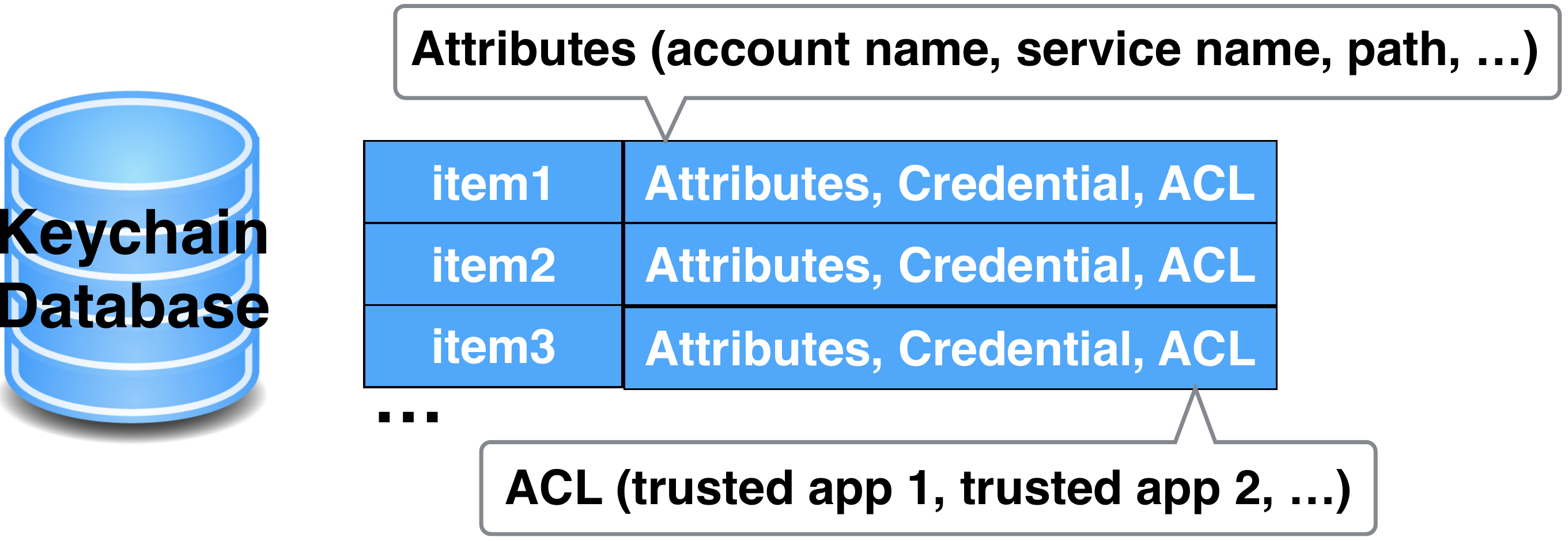}
\vspace{-10pt}
\vspace{-2pt}
\caption{Simplified Keychain Structure}
\label{fig:keychain}

\vspace{-10pt}
\end{figure}

\begin{figure}[h]
\centering
\includegraphics[scale=0.35]{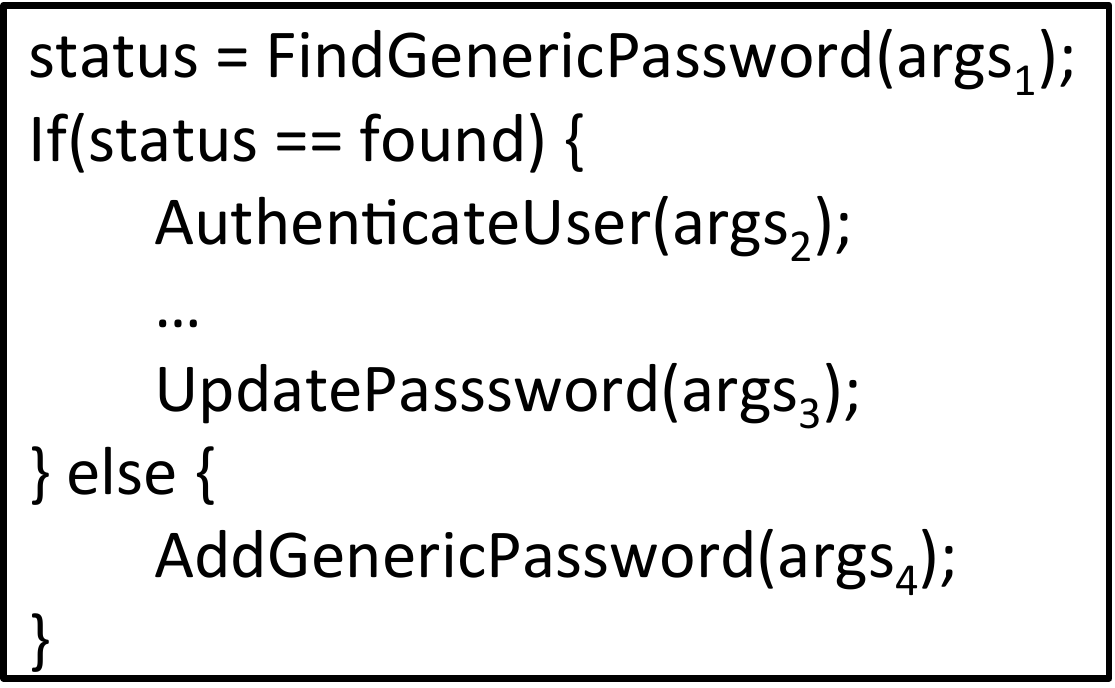}
\vspace{-10pt}
\vspace{-2pt}
\caption{Workflow of Keychain Template Code by Apple}
\label{fig:keychainsample}

\end{figure}

\vspace {3pt}\noindent\textbf{Security weakness}. The simplified keychain structure is illustrated in Figure~\ref{fig:keychain}. Each keychain item carries the credential (e.g., password, secret key, etc.) under protection and a set of attributes, such as account name, service name, path, etc. Types of attributes an item has depend on its class, typically \textit{Internet passwords} or \textit{generic passwords}.  Figure~\ref{fig:keychainsample} further shows how the keychain should be used according to Apple~\cite{templatecode}. An app first searches the keychain using a set of attributes to find out whether its item has already been there\footnote{\small This happens, for example, when the app has been upgraded from a lower version already in the system.}\ignore{, through the API \texttt{SecKeychainFindInternetPasswo\\rd} or \texttt{SecKeychainFindGenericPassword}}. If so, the item  should be \ignore{is automatically}updated to keep the app's current credential, after the app has been authenticated (signature verification) and authorized (ACL lookup) by the OS. Otherwise, the app \ignore{calls \texttt{SecKeychainA\\ddInternetPassword} or \texttt{SecKeychainAddGenericPass\\word} to create}creates a new item and set attributes to index it.

On OS~X, the creator of a keychain item can also attach to it an access control list\ignore{, using the function \texttt{SecAccessCreate}}. The ACL includes the operations that can be performed on the item (e.g., read, write, etc.) and a set of trusted apps with the permissions to do so. Whenever an app attempts to access an item, the service first checks whether the access is allowed to happen and denies it when it is not. Then, the service further looks up the ACL: when the app is not there, the user's permission is required to let the operation proceed.

With its careful design, this access-control mechanism was found in our research to still contain security-critical vulnerabilities, allowing a malicious app to hijack a target app's keychain item. One scenario for this exploit is that when the malware runs before the victim app creates a password (or rather a keychain item) in the keychain. What the attacker can do here is to use the attributes of the target app (the victim) to claim an item and also craft an ACL that includes the target as a trusted app. When the target uses the keychain to store password, it discovers the item with its attributes already there and treats the item as its own secure storage (illustrated by the Apple's template code in Figure~\ref{fig:keychainsample}). Note that this is reasonable given that an app's older version or other apps from the same developer may have already been installed on the system. Since the target is on the ACL of the item (which is controlled by the attacker), the OS allows all its operations to proceed. Therefore, at no point the target gets any indications from the keychain that it is just a guest user of the item, and the owner is untrusted. This confusion will cause the target to divulge its secrets to the attacker, whenever it updates the user's credentials to the keychain. 

Apparently, the attack can only succeed when the attributes of the victim's keychain item are predictable. This is mostly the case and the attributes typically remain constant for specific apps or services\ignore{, except the \texttt{account name} part which depends on the specific victim (usually the user's email address)}.  Also, the attacker needs to create the keychain item first. These restrictions, however, turn out to be unnecessary: we found that the attributes of \textit{any} keychain item are actually \textit{public}, though their content (credential) is protected. And most importantly, we found a second flaw in keychain that an existing keychain item can be deleted by an unauthorized sandboxed app.  As a result, all the attacker needs to do is just identifying an existing item, removing it from the keychain and creating a new one of its own with the same attributes to wait for the target app to put its secret there. Fundamentally, the problem comes from the challenge for an app to authenticate the owner of an existing keychain item. Apple does not offer a convenient way to do so.  Little information is given to an app even for identifying the owner of an item, not to mention any authentication support.  The only way that could mitigate the threat is for the target app to inspect an existing item's ACL, making sure that its name is not on the list together with untrusted parties. However, Apple has never mentioned that this should be done. As a result, protection is not in place within apps, leaving them completely vulnerable to our attacks.



A straightforward solution is to strip some functionalities from the keychain, making it simple. Actually, iOS does not have this issue, because its keychain does not support the ACL at all: every app is only allowed to access its own item and there is no flexibility to let a group of apps share secrets except those by same developers. This works because iOS apps are pretty simple and do not need much collaboration, which is not the case on OS~X.  For example, Safari manages the user's passwords for different websites stored by other browsers such as Chrome, which is made possible through the keychain's access control mechanism. Also, given the trend that the iOS apps become increasingly complicated, the demand for such collaboration may show up in the future.

\vspace {3pt}\noindent\textbf{Attacks}. In our study, we utilized an automatic tool to identify hundreds of vulnerable apps (Section~\ref{sec:detectandmeasure}). Here we elaborate our end-to-end attacks on two prominent examples, Apple's \textit{Internet Accounts} and Google Chrome. Internet Accounts is an OS~X system app that manages the user's various Internet accounts, e.g., those for iCloud, Email, Twitter, Facebook and others.  The app stores those accounts' secret tokens in the default keychain, each in a different item. In a similar way, Chrome also keeps the user's passwords for each web account.


In our research, we built a sandboxed attack app against Apple's \textit{Internet Accounts} on OS~X 10.10, the most recent version when we found the problem, and Chrome v40.0.2214.94. The attack app managed to hijack the keychain item the Internet Accounts app uses for keeping the iCloud token and the item in which Chrome stores the user's Facebook password. Note that all other information these apps put there is equally vulnerable to the same attack.  Specifically, iCloud utilizes the user's email address as her account name and sets the \texttt{serviceName} attribute to ``Apple ID Authentication''.  Other attributes the Internet Accounts needs to create and retrieve a keychain item are just the string length of these two attributes (\texttt{accoutName} and \texttt{serviceName}). Our attack app, running before the service was set up, first created an item using these attributes and deliberately granted the full access permission to the Internet Accounts.  As a result, the Apple service unwittingly updated to the item the user's iCloud token\ignore{, after it was configured by the user}. In a similar way, the tokens of Gmail, Facebook, Twitter, etc. managed by Internet Accounts are also exposed to the attack app.

When it comes to Chrome,  the browser also utilizes attributes \texttt{serviceName} (\url{www.facebook.com}), \texttt{accoutName} (email address of the user) and their lengths. Other attributes involved are the URL path (a constant `/'), port (always `0'), protocol (\texttt{kSecPro\\tocolTypeHTTPS}) and authentication type (\texttt{kSecAuthentic\\ationTypeHTMLForm}).\ignore{ All these attributes are predictable, allowing the attacker to pre-empt the keychain item.  As a result,} Our app successfully hijacked the keychain item and obtained the user's Facebook password. It further got through the MAC App Store's security checks. Video demos for both attacks are posted online~\cite{supporting}.

We reported this vulnerability to Apple on Oct. 15, 2014, and communicated with them again in November, 2014 and early 2015. They informed us that given the nature of the problem, they need 6 months to fix it. We checked the most recent OS~X 10.10.3 and beta version 10.10.4 and found that they attempted to address the iCloud issue using a 9-digit random number as \texttt{accountName}.\ignore{ The number is obtained by the app from the Apple server to identify an iCloud user.} However, the \texttt{accountName} attribute for other services, e.g. Gmail, are still the user's email address. Most importantly, such protection, based upon a secret attribute name, does not work when the attacker reads the attribute names of an existing item and then deletes it to create a clone under its control, a new problem we discovered after the first keychain vulnerability report and are helping Apple fix it.




\subsection{Container Cracking}
\label{subsec:bundleid}


The security weaknesses within the keychain happen when sandboxed apps want to share resources (i.e., passwords) across sandbox boundary. However, even for the private resources inside each app's sandbox which are never designed for sharing, XARA attacks can still happen, due to a weakness in the unique BID-based separation design on OS~X.

\vspace {3pt}\noindent\textbf{BID conflict}. As introduced in Section~\ref{sec:background}, each sandboxed app has a BID, which needs to be unique.  This is important because once the app is installed, its BID is used to create a container directory that other sandboxed apps cannot touch. On OS~X, all apps' containers are under the directory $\sim$\url{/Library/Containers/}, e.g., $\sim$\url{/Library/Containers/com.evernote.Evernote/}.  \\Their directory names, the BIDs, bind them to their individual apps: the OS verifies app signatures whenever access attempts are made, and only those from the owners of the directories or the parties on their ACLs are allowed to go through.  To ensure the uniqueness of BIDs, the MAC App Store checks submissions to deny those using the BIDs of the apps already in the store\footnote{\small Note that Gatekeeper typically blocks the installation of untrusted third-party apps.}.

What causes a complication here is the embedded programs within an app, that is, the \textit{sub-targets} of the app's project. A sub-target can be a helper program, XPC Service (another MAC IPC mechanism), or framework, etc. each of which has its own plist and BID. Particularly, the helper (e.g. 1Password mini program) and the XPC Service also have their individual container directories, while the framework is a directory for shared resources (e.g., libraries). For apps published by the Apple Stores, their helpers and XPC Services are all sandboxed. Once installed, their containers are also placed under \texttt{$\sim$/Library/Containers/}, alongside with those of their main programs. Interestingly, we found in our research that the MAC Store fails to verify whether a sub-target's BID is in conflict with those belonging to other apps or their sub-targets, except for the Apple reserved BID (those starting with \url{com.apple}). This allows one to publish an attack app whose helpers or XPC Services are using the BIDs of other apps, their helpers or XPC Services.  Once the attack app is launched, whenever the OS finds out that the container directory bearing the sub-target's BID (as its name) already exists, the sub-target is automatically added onto the directory's ACL.  As a result, the malicious app gains the full access to other apps' containers, which completely breaks its sandbox confinement.


The cause of the problem could be the convenience given to the app developer to share frameworks, helpers or XPC Services in different apps. Particularly, in our study, we scanned 1,612 apps from the Mac App Store and \textit{found 40 frameworks shared by different developers, e.g., \texttt{DropboxOSX.framework} used by 14 apps for subscribing the Dropbox service}. This security risk is not present on iOS, on which the containers of main programs and sub-targets are put under different parent directories, and most importantly, they are named with randomly generated UUIDs. Again, the simplicity of the container structures here could be the result of limited functionalities of iOS apps, which do not need to extensively share resources among them.


\vspace {3pt}\noindent\textbf{Attack}. This BID conflict threat affects every sandboxed app running on OS~X. In our study,  we implemented end-to-end attacks on a few high-profile apps, including Evernote, WeChat, QQ (a popular online chat app), Money Control (a popular Finance app) and others (Section~\ref{subsec:impact}). For example, from the container of Evernote, our attack app, involving an XPC Service that hijacked the target app's BID, successfully stole all the contacts of the user and her private notes from \texttt{$\sim$/Library/Containers/com.evernote.\\Evernote/account/}.  Also, it recovered all the message photos under WeChat and QQ. Again, our app got through the security check of the MAC App Store. The video demos of the attacks can be found at ~\cite{supporting}. In Section~\ref{subsec:impact}, we further present the consequences of the attacks on other apps.



\subsection{IPC Interception}
\label{subsec:IPC}


Breaches of cross-app resource sharing (i.e., keychain) and BID based sandbox isolation mechanism unwittingly grant the adversary unauthorized access to other apps' resources. The problem, unfortunately, does not stop here: in our research, we found that major cross-app communication (IPC) channels on OS~X, NSConnection in particular, and those deployed across platforms, such as WebSocket and Scheme (Section~\ref{subsec:scheme}), are also designed with flaws. This exposes critical information, e.g. all Web passwords in major browsers, to the adversary in even more various ways.  Below we elaborate our findings\ignore{ about NSConnection, XPC and WebSocket, and leave URL scheme, a more popular channel for both OS~X and iOS}.

\vspace {3pt}\noindent\textbf{NSConnection}. As introduced before, NSConnection is an Apple-specific IPC channel. It allows one party to act as the server and share an object with other client apps.  These clients can then communicate with the server by invoking the interfaces defined within the object.  The channel is designed to deliver a large amount of data between apps, compared with Scheme (Section~\ref{subsec:scheme}). A security problem here is that the OS does \textit{not} offer means for the apps to authenticate each other when they are using NSConnection, nor does Apple ask the app developer to avoid sending secrets across the channel. As a result, we found that a sandboxed app can easily impersonate the server or the client of a target app to the other party to access sensitive resources.


Specifically, to create an NSConnection object, the server needs to have a name for the object, which is typically a constant string (e.g., \texttt{com.evernote.ipc.client} for Evernote) hardcoded within the app.  This name is later used by the client to acquire the object from the OS (through the API \texttt{rootProxyForConnecti\\onWithRegisteredName}). What can happen here is that an attack app can create an NSConnection object with that name, ahead of the target server, to impersonate it to the client.  In this case, the client will be cheated into communicating with the attacker, taking it as the target. The attack can also go the other way around: the malicious app, with the knowledge of the target server's name, can connect to it and use the interfaces of its object to invoke its internal functions. Note that Apple does not offer any API to let the server or the client find out the identity of the party it is talking to (e.g., the process ID of the app).  Therefore, authentication in the IPC is not supported by the OS. Given that the need for such authentication has never been made clear by Apple, all apps using this channel were found to be vulnerable in our research (Section~\ref{sec:detectandmeasure}).


\vspace {3pt}\noindent\textbf{WebSocket and beyond}. Unlike NSConnection, WebSocket is not Apple-specific, and instead a generic protocol for a server and a client to establish a full-duplex single socket connection. Its specification~\cite{WebSocket}, which has been developed as part of the HTML5 standard, introduces a JavaScript interface through which the web content inside a browser or an app's webview instance can directly talk to another app. This channel is often used by browser extensions to communicate with an app on the local system through a predetermined TCP port. Specifically, the app runs a WebSocket server to listen on the port, which is connected by the script code of the extension to exchange data. The problem is that in the absence of proper authentication, a malicious program (with the network permission when it is sandboxed) can preemptively claim the port before the legitimate server does. This enables it to receive data from the target extension.  Such a security risk can also happen on the browser side: a malicious extension can impersonate the authorized one to talk to the local app through its port. Note that other inter-app communication through TCP port, like the local web server used by popular app Pushbullet, can also be attacked in this way.

It turns out that the Apple platforms do not provide any means for an extension to authenticate a local WebSocket server. There is no way for the extension to find out the identity of the local app through an API call.  The only solution is a custom authentication mechanism built by the app/extension developer.  On the other hand, major browsers, e.g. Google Chrome, embed the ID of an extension in their message delivered to the local program, which helps the latter to determine whether the message comes from the right party.\ignore{ Note that this is not a common feature.  Other browsers, such as Firefox, do not provide such information.}  However, since the extension impersonation threat has not been identified, the developer has not been informed about the importance of building proper protection into her app.

\vspace {3pt}\noindent\textbf{Attacks}. The security risks of intercepting the IPC communication through these vulnerable channels are realistic and serious. As an example, here we just elaborate our end-to-end attacks on three popular apps.\ignore{ More information is available in Section~\ref{subsec:impact}.} We analyzed the 1Password app for OS~X, which is one of the most popular password management apps and ranked 3rd by the MAC App Store~\cite{1pass}. The app comes with a browser extension for each major browser that collects the passwords from the user's web account and passes them to the app through a WebSocket connection. In our research, our sandboxed app created a local WebSocket server that took over the port 6263, before the 1Password app did, and was successfully connected to the password extension and downloaded the password whenever the user logged into her web account.\ignore{ Also, our attack extension with a permission to visit only a specific domain was able to steal the passwords for other domains, including Facebook, Google, etc., from the 1Password app through a WebSocket connection to its port.}  We reported our findings to the 1Password security team, which acknowledged the gravity of this problem.  This attack succeeded on OS~X 10.10 (latest version when we reported the problem), against Chrome, Firefox and Safari. Our attack code passed the vetting process of the MAC Store. The attack demo is here~\cite{supporting}.

The similar attack was also successful on Pushbullet, an Apple-recommended popular app for exchanging notes, links, pictures and files between multiple devices. The app authenticates a user by running Google Single Sign On (SSO) within a browser. After the user signs in, Google redirects the browser to the app's local web server that listens on the port 20807, together with a secret token.  In our attack, this port was first taken over by the malicious app, which then got the redirection link from Google and stole the token. After that, the attacker released the port to Pushbullet, which later got the token resent by Google.

We further exploited the NSConnection channel used by the famous Evernote app, the most popular notetaking and archiving app on the Apple platforms (ranked 3rd among free ``Productivity'' apps in the MAC Store).  Evernote includes an NSConnection server to exchange data with its helper program.  What we found is that an attack app can impersonate Evernote's server before it starts to run and communicate with the helper.  Also, it can act as the NSConnection client to get an object from the server.  The object our app obtained allowed the attacker to acquire the authentication token of the Evernote app.  Our demo is posted here~\cite{supporting}. 

\subsection{Scheme Hijacking}
\label{subsec:scheme}

As mentioned earlier, URL Scheme, an inter-app communication channel, is different on the Apple platforms. Specifically, Apple's OSes automatically associate a scheme with one app even with the presence of multiple apps claiming the same scheme. This design leads to a unique problem to Apple's OSes, \ignore{Compared with NSConnection and WebSocket, Scheme is simpler and does not involve server and client components. However, this channel is equally vulnerable} \ignore{: our research shows that oftentimes, Scheme has not been used by app developers in a secure manner, which opens an avenue for an unauthorized app,  even when it is confined within the sandbox, to access other apps' internal resources}as elaborated below.  

\vspace {3pt}\noindent\textbf{Scheme takeover}.  Essentially, a URL scheme is a simple protocol that an app defines for communicating with others. The app specifies a URL format in its plist file and lets other apps invoke it and pass parameters through the URL. Once this URL is triggered within the browser or a webview instance inside another app, an HTTP redirection is launched towards the ``location'' part of the URL, e.g., ``\texttt{yelp:}'', and thus activates the app claiming the scheme, using the data delivered by the remaining part of the URL.  Also, a scheme invocation can be initiated by an app with the API \texttt{openURL}. Apple extensively utilizes URL schemes to run system apps, e.g., \texttt{mailto} (for opening the Mail app), \texttt{tel}, \texttt{facetime} and \texttt{sms} (for launching their corresponding apps).


Things become a bit tricky when two different apps register the same URL scheme with the OS.  This conflict is resolved on the Apple platforms according to the nature of the scheme. Specifically, Apple has a list of system schemes (e.g., \texttt{sms}, \texttt{Facetime}, etc.) that cannot be taken by any third-party apps, and another list of schemes whose affiliations can be changed under the user's consent, e.g., the default browser for opening \texttt{http}. For a scheme not on the lists, \textit{it will typically be bound to the first app that registers it on OS~X and the last on iOS}, as discovered in our study. Given this conflict resolving strategy, a malicious program can hijack a target app's scheme to get the service request or even the data sent to it. Particularly on iOS, the attack works even when the malware is installed after the target app.



This scheme hijacking attack can be detected on OS~X using the API \texttt{URLForApplicationToOpenURL} or \texttt{LSCopyDefault\\HandlerForURLScheme}, which returns the identity of the app that successfully registers a given scheme. However, no corresponding API exists on the other Apple's OS, i.e., iOS. \textit{In the absence of such OS-level supports, an app can do nothing to authenticate the party it invokes through a URL}.  Therefore, all third-party apps running on iPhone and iPad are completely unprotected from this threat.  Note that Apple has never explicitly asked its developers to verify the apps launched by URLs, nor does it check duplicated scheme definitions at the App Stores, as observed in our study.  The consequence is that oftentimes, even OS~X apps are less protected than they should, and vulnerable to the scheme hijacking attack (Section~\ref{subsec:impact}).  Following we elaborated our end-to-end attacks on some high-profile OS~X and iOS apps. Note that our attack apps were successfully uploaded to both the MAC and iOS App Stores.

\vspace {3pt}\noindent\textbf{Attack on OS~X}. In our research, we implemented an attack on Wunderlist, a popular free app (ranked 5th in the ``Productivity'' category on the MAC Store) for managing MAC users' to-do lists. The app uses Google SSO: the user logs in Google in the browser and then is redirected to the URL \texttt{wunderlist://oauth/goo\\gle?token=ya29XXX}, which invokes Wunderlist, passing to it a secret token.  In our attack, an unauthorized app first registered the scheme ``\url{wunderlist://}''. As a result, our app stole the token from the browser. More interestingly, our malicious app then immediately delivered the token to Wunderlist by calling \url{openURLs:withAppBundleID} (an OS~X specific API), acting as a man-in-the-middle. This actually lets the SSO go through and therefore make the attack stealthy. (The attack demo here~\cite{supporting}.)


\vspace {3pt}\noindent\textbf{Attack on iOS}. Scheme hijacking poses an especially serious threat to iOS, which heavily relies on URLs for inter-app interactions (Section~\ref{subsec:impact}).  As an example, we exploited this weakness and successfully obtained the victim's Facebook token for Pinterest, the most famous personal media management app. Specifically, Pinterest and other apps all support the SSO login through the Facebook app. Whenever the user clicks on ``continue with Facebook'' in these apps, the Facebook app is invoked to ask for the user's permission to let the authentication go through and also grant Pinterest (and other apps) access to some of her Facebook data.  With the user's consent, Facebook triggers a scheme \url{fb274266067164://access_token=CAAAAP9uIENwBAKk&X=Y} to deliver a secret access token to the app. In our research, our attack app registered ``\url{fb274266067164://}'' and took over this scheme. As a result, Facebook unwittingly launched our app and passed to it Pinterest's access token. Actually even the scheme for invoking the Facebook app (``\texttt{fbauth://}'') was successfully hijacked in another attack, which enabled the attacker to become a man-in-the-middle, performing the whole SSO within its webview instance on behalf of the real Facebook app. Most importantly, once it got the secret token from Facebook, the attacker forwarded it to the Pinterest app through ``\url{fb274266067164://}'', which completely hid the attack from the user. Note that this last step can be detected using the API \texttt{openURL:sourceApplication}, which returns to the caller (here, Pinterest) the party that initiates the scheme communication. However, the protection is not in place within any apps that we scanned (Section~\ref{subsec:impact}), including Pinterest. This may be due to the fact that Apple never explicitly informs the developers to do this inspection. Interestingly, again there's no corresponding API for detection on the other Apple's OS, i.e., OS~X this time. We successfully launched the attack. (The attack demo here~\cite{supporting}.)

\vspace {-5pt}
\section{Measurement and Defense}
\label{sec:detectandmeasure}

\ignore{Our security analysis on the Apple platforms show that their isolation protection (including sandboxing and other security policies enforced within the OS, and the security checks performed by the Apple Stores) can be circumvented by the knowledgeable adversary.  Still less clear here are the impacts of such security weaknesses, which depend on how likely real-world apps can be exploited through the weaknesses and what the consequences will be. Also important here is the effort that can be made to mitigate this new threat.  }In this section, we elaborate an automatic analysis tool we built for detecting vulnerable apps and a measurement study that reveals the scope and magnitude of the XARA problem.  We further show that though a generic solution to the problem needs a significant effort from Apple and its app developers, a simple native program operating on MAC OS~X can help mitigate the threats.

\subsection{Detection of Vulnerable Apps}
\label{subsec:detect}

Among all the security weaknesses reported in Section~\ref{sec:attack}, some (e.g., the scheme hijacking on iOS, the BID conflict, NSConnection) are caused entirely by the security flaws within the system (OS or the Apple Store), and only a system-level solution can fix them. Other threats, however, are more contingent upon what an app does, particularly, whether it properly authenticates the party it interacts with.  To better understand the impacts of those security weaknesses, we developed \textit{Xavus}, an XARA vulnerability scanner that statically inspects Apple apps' binaries to identify those susceptible to the XARA threat\ignore{, that is, the ones utilizing various less protected channels (Section~\ref{sec:attack}) without proper authentication in place}. This analyzer was then used to evaluate the security qualities of a set of popular apps\ignore{ on the Apple platforms} (Section~\ref{subsec:impact}). Note that Xavus can also serve the developer by helping her identify XARA weaknesses in her app, which is important given the challenges in fixing the problems on the system level: e.g., the keychain issue we reported last October is still not successfully fixed in the most recent OS~X 10.10.3 and beta version 10.10.4.

\begin{figure}[h]
\centering
\vspace{-7pt}
\includegraphics[width=0.9\linewidth]{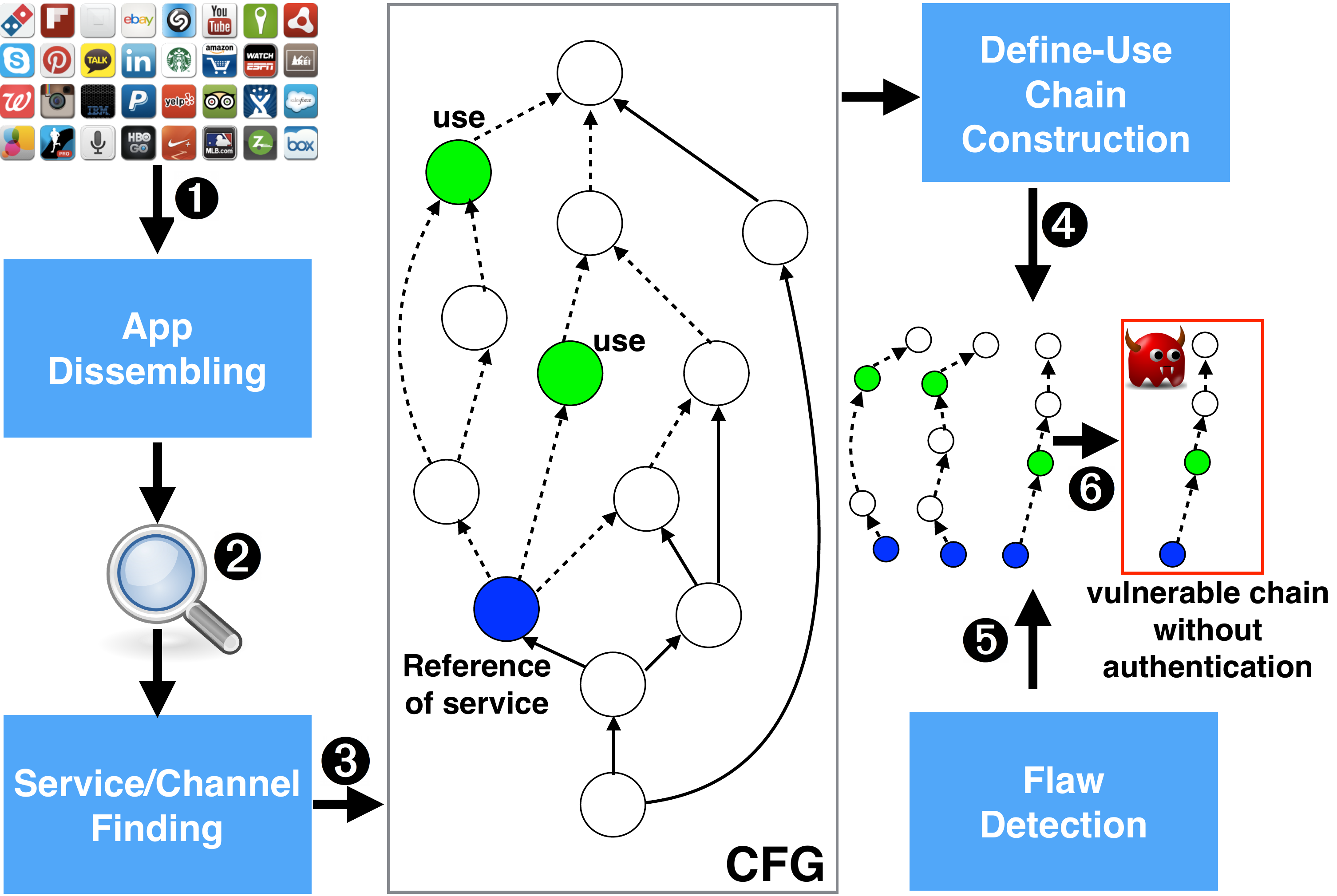}
\vspace{-10pt}
\vspace{-2pt}
\caption{Detection of XARA Vulnerability}
\label{fig:detection}

\vspace{-7pt}
\end{figure}

\vspace {3pt}\noindent\textbf{Design}. The idea for detecting the XARA vulnerability within an app is to find out whether the app authenticates other parties associated with a service (e.g., keychain) or a channel (e.g., WebSocket, NSConnection or Scheme) before using it.  Since typically, one needs to first \textit{claim} such a service or channel, what we need to do is to inspect the control-flow graph (CFG) between the program location for the claim and that for the use and find out whether the authentication has happened. To this end, Xavus is designed to include five modules, as illustrated in  Figure~\ref{fig:detection}, for disassembling an app's binary, determining whether a specific service or channel is utilized, and if so constructing CFG and define-use chains\ignore{and if so the locations of the claim and the use,}, and identifying the presence of authentication on the define-use chains\ignore{program execution paths connecting these two locations}. Following we describe how this design works and how it was implemented in our research.

\vspace {-3pt}
\vspace {3pt}\noindent\textbf{Apple code analysis}. The first step for analyzing an MAC or iOS app is to disassemble its binary.  Most Apple's apps are built with Objective-C and their binaries are in the Mach-O format.  Disassembling such binaries is done within our implementation of Xavus through \textit{Hopper}~\cite{hopper}, a popular tool for reverse-engineering Mach-O files~\cite{MachO}, which converts the MAC binaries into Intel x64 instructions and iOS app code into ARMv7 instructions under our settings.  Note that iOS apps are encrypted and need to be decrypted before they can be analyzed by Xavus.  In our research, we ran Clutch~\cite{Clutch} to decrypt iOS apps \ignore{a virtual-machine based memory dumper to}and then collect their binaries.

To determine the type of services or channels an app uses, our analyzer inspects its instructions for related API calls, e.g., \texttt{SecKeyc\\hainFindGenericPassword} for keychain access. The names of these functions are all kept under a specific section of the app's Mach-O file.  Such a function is invoked by the Objective-C binary in a unique way: it is\ignore{ not called directly or using virtual method tables, and instead,} triggered by sending messages to the function's object, which happens through passing a pointer to the message receiver, the name of the function (called \textit{selector}, a null-terminated string), together with other parameters to the runtime function \texttt{objc\_msgSend}. This operation can be observed from an app's recovered instruction set when the selector (i.e., function name) is stored to the RSI register for preparing the \texttt{objc\_msgSend} call. Leveraging this observation, Xavus can find out the program location where a service or channel is claimed and where it is used.

\ignore{In order to detect vulnerable apps that use resources without authentication, we need not only to search for the existence of authentication API, but also to make sure whether the authentication actually happens on the servic or channel that apps claimed. To achive that goal, we need to do static analysis to detect whether the service or channel that the app claimed is passed to any authentication API along its define-use chain between the claiming and using of the channel.}

\begin{figure}[h]
\centering
\vspace{-3pt}
\includegraphics[width=0.99\linewidth]{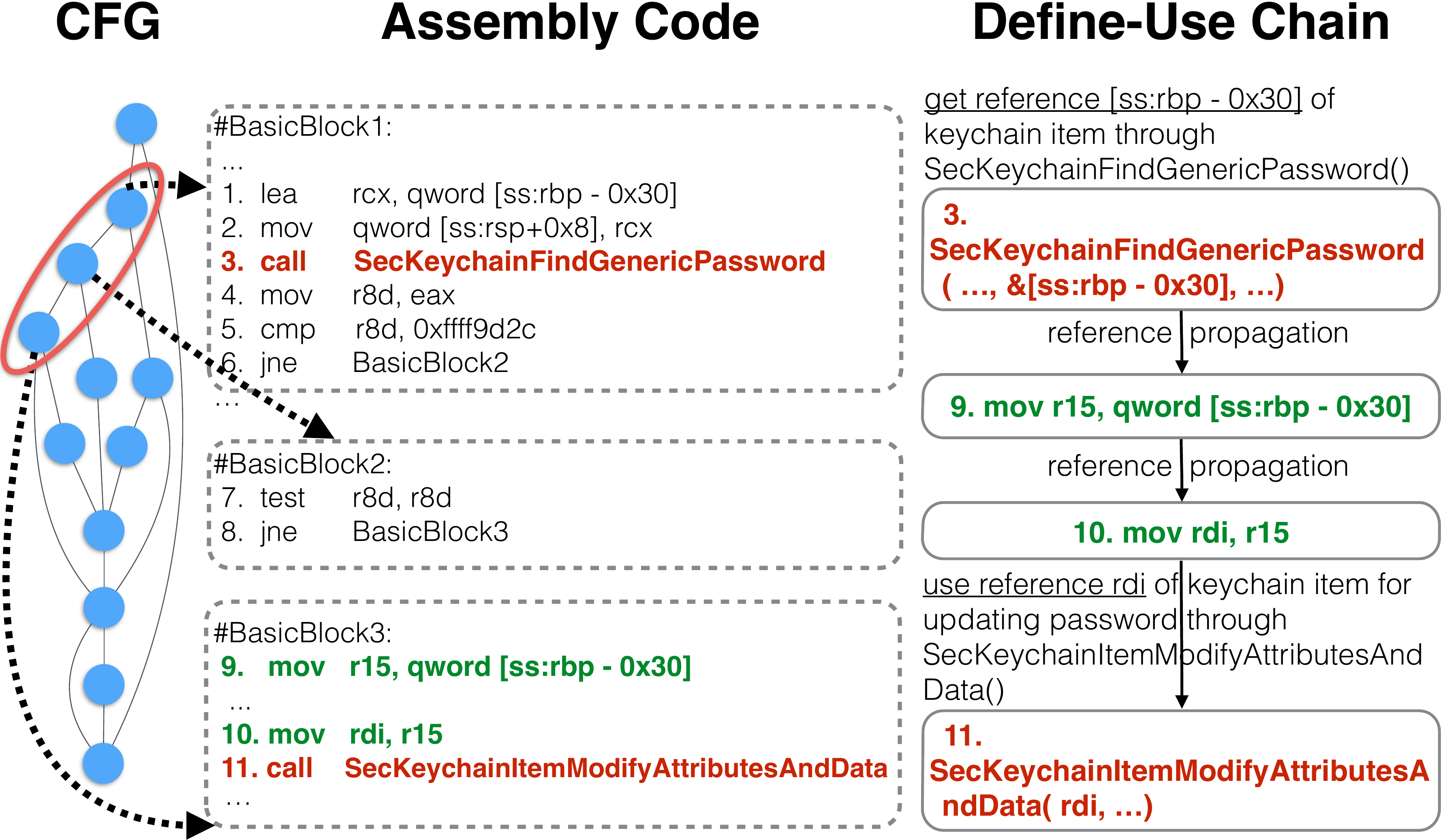}
\vspace{-10pt}
\caption{Detection of Evernote's Vulnerable Code for Keychain Password Updating}
\label{fig:assembly_sample}

\vspace{-4pt}
\end{figure}

To detect vulnerable apps that claim and use services or channels without authentication, we need to not only search for the presence of authentication API, but also check whether the authentication actually happens to the services or channels in use. To this end, our approach first searches for the invocation of the API for claiming a service or channel, and runs Hopper scripts to construct a CFG for the procedure involving that API call (Figure~\ref{fig:detection}). Then, from the location of the claim, Xavus performs a define-use analysis: from the \textit{reference} of the service/channel (e.g., a pointer to a keychain item, or the constant string for a scheme), as returned by the API for the resource claim\ignore{(e.g., assembly code line 3 in Figure~\ref{fig:assembly_sample})}, our approach identifies all the program locations where the reference is utilized before it has been redefined (e.g., the variable holding the reference is assigned with a different value). The objective here is to locate another API for using the service or channel (e.g., updating a password to a keychain), so as to find out whether proper authentication happens before the use.  Typically, the claim (getting the reference) for a service/channel and the use of the service/channel (through the reference) stay in the same procedure. Occasionally, these API calls are placed in different functions, and need to be linked by constructing an inter-procedure CFG. This can be done using the techniques described in the prior research~\cite{egele11:pios}. Along the execution path identified during the construction of the define-use chain,  our analysis module further checks whether the reference of the service/channel is passed to any authentication operation, according to the types of the services and channels in use. \ignore{ Along the execution path identified during the construction of the define-use chain,  our analysis module further checks the presence of authentication operations which are associated with the reference of the service/channel, according to the types of the services and channels in use.}

Figure~\ref{fig:assembly_sample} shows an example for the analysis on Mac Evernote app (Version 6.0.9), which does not authenticate the owner of a keychain item before updating to it the user's password. Specifically, Xavus first builds the CFG of an Evernote procedure \texttt{[ENKeychai\\nHelper saveValue:toKeyChainItem]}, and locates the claim for the keychain service, i.e., getting a reference of an item through the API \texttt{SecKeychainFindGenericPassword}\\ (Line 3 under Assembly Code in Figure~\ref{fig:assembly_sample}). The reference is returned in memory \texttt{[ss:rbp - 0x30]}. Then Xavus follows the propagation of the reference and identifies one instance of using it, i.e., updating passwords through the OS~X API \texttt{SecKeychainIt\\emModifyAttributesAndData}. Xavus reports that Evernote is vulnerable due to the absence of authentication before the use of the reference.

\vspace {3pt}\noindent\textbf{Flaw detection}. As mentioned earlier, Xavus is designed to detect missing authentication of the parties that share a service or a channel with an app before it is utilized by the app, a \textit{necessary} condition for the presence of the XARA weakness. To this end, we built into our analyzer a set of features to fingerprint the authentication operation for each channel or service, as elaborated below.

\vspace {3pt}\noindent$\bullet$\textit{ Keychain}. The claim of this service happens when the app calls \texttt{SecKeychainFindGenericPassword} or \texttt{SecKeychainF\\indInternetPassword}. Both APIs return \texttt{itemRef}, the reference to the item.  This reference will ultimately be used to update a password to the keychain item through either \texttt{SecKeychainIt\\emModifyAttributesAndData} or \texttt{SecKeychainItemMo\\difyContent}. The only possible way that app could find out whether the item is created by a trusted party is to inspect its ACL that documents all the apps allowed to access the item. This is done through the API \texttt{SecACLCopyContents}.  The absence of a call to the API on the execution path (between the claim and the use locations) indicates the existence of a XARA risk within the app.

\vspace {3pt}\noindent$\bullet$\textit{ NSConnection}. As discussed in Section~\ref{subsec:IPC}, NSConnection involves a server and a client. On the client side, the app's claim for the channel (connecting to the server) is identified from the API \texttt{rootProxyForConnectionWithRegisteredName} or \texttt{co\\nnectionWithRegisteredName}, which returns an object fr-\\om the server. Whenever any parameter on the object is found to be passed to a function call, we consider that the channel is used.  Between these ends, authentication needs to be done for the client to verify the identity of the server sharing the \texttt{NSConnection} object.  This, however, is \textit{not} supported on today's MAC OS~X. More specifically, the name of the object is mapped within the OS to an \texttt{NSMachPort} object directly related to the server app.  The problem is that according to Apple, verification of an app through its \texttt{NSMachPort} has yet been implemented~\cite{machnotimp}. Note that once this support is provided, we can look for the API \texttt{SecCodeCheckVal\\idity} to determine whether the authentication is in place. On the server side, whenever the app calls \texttt{serviceConnectionWith\\Name} to share its object, it loses the control on the object. Any party can get it from the OS.


\vspace {3pt}\noindent$\bullet$\textit{ WebSocket}. WebSocket servers are typically built over a few popular open-source frameworks, such as CocoaHTTPServer~\cite{CocoaHTTPServer} and QtWebKit~\cite{QtWebKit}. All of them provide a \texttt{receiver} method for getting messages from browser extensions, which is used by Xavus to fingerprint this channel, and a \texttt{response} method for replying to the extensions. The invocations of these two methods are identified as the claim and the use of the channel, respectively. Between them, the server is supposed to access the HTTP header \texttt{Origin} that includes extension IDs attached by the browser and check the signature of the browser through the API \texttt{SecCodeCheckValidity}. If these operations are not found, the app is considered vulnerable to the threat from a malicious extension or app.  On the other hand, the attack from a malicious server against an extension cannot be detected through any existing APIs.



\vspace {3pt}\noindent$\bullet$\textit{ Scheme}. The most popular ways for one app to launch another through the Scheme channel is to trigger a constant link embedded in the app or a URL returned from its webview instance. The former can be directly found from the app, while the latter comes from one of the four methods of \texttt{WebPolicyDelegate}, an object that lets an app control the operations on the web content within the webview. On the list are \texttt{decidePolicyForMIMEType:reques\\t:}, \texttt{decidePolicyForNavigationAction:request:},  \texttt{d\\ecidePolicyForNewWindowAction:request:} and \texttt{wil\\lPerformClientRedirectToURL:}. The program location where any of these methods is invoked or the constant string is accessed is considered to be a claim for the Scheme channel.  The use of the channel happens when the URL is invoked through \texttt{openURL}.

On OS~X, between the claim and the use, the app can run \texttt{URLFor\\ApplicationToOpenURL} or \texttt{LSCopyDefaultHandlerFor\\URLScheme} to find out which app will be launched by a given scheme. In the absence of any of these calls, it is highly likely that the app is vulnerable to the scheme hijacking attack. An exception is when the scheme here is actually claimed by the OS, for example, \texttt{mail}, \texttt{facetime}, etc., which can be easily identified when the URL is a constant string within an app. In the case that it actually comes from the web, the chance is that it is indeed vulnerable, as the webview can be used to open any links. Note that for iOS, an app does not have any means to find out the owner of a scheme.

\vspace {3pt}\noindent$\bullet$\textit{ BID}. The BID confusion problem is completely caused by the Apple Store and the design of the MAC OS~X sandbox. Nevertheless, Xavus is built to find out whether apps deposit data to their containers, which indicates that sensitive information could be exposed through this vulnerability.  This data-storing activity can be easily identified from the function call \texttt{NSHomeDirectory}.

\vspace {3pt}\noindent\textbf{Discussion}. Our current implementation of Xavus works effectively on most apps over the Apple platforms, as observed in our research (Section~\ref{subsec:impact}).  \ignore{Xavus can not only inspect all the cross-app interactions channels discussed in this research, its design also make it scalable to detect missing authentications in other cross-app channels once the features to fingerprint the authentications for that channel are established.}Even though Xavus was implemented for detecting the XARA on known channels,  the idea behind it,  authentication check between the claim and the use of a channel,  could find its application to detecting similar flaws within other cross-app mechanisms. However, there are situations when a vulnerable app falls through the crack: for example, when the app dynamically generates the scheme to be invoked.  On the other hand, the app developer might implement some ad-hoc protection that our analyzer misses. This happens to those using the keychain or WebSocket. An app could delete its keychain item and create a new one each time when it updates new credentials there, an approach that Apple does not recommend~\cite{kcitemdelete}.  \ignore{An app can choose a secret attribute for its keychain item, making it unpredictable to the adversary, as did by Apple to protect its Internet Accounts, in response to our finding.} Also, a browser extension may authenticate a local program using a secret over the WebSocket channel.  In our research, we manually analyzed the apps randomly sampled from those flagged as vulnerable by our implementation to ensure that the results were accurate, and concluded that they indeed were in a vast majority of cases.

\vspace {-5pt}
\subsection{Impacts}
\label{subsec:impact}

With the help of Xavus, we were able to analyze a large number of popular Apple apps to understand their susceptibility to the XARA threat. In our study, we downloaded 1,612 free apps from the MAC App Store. These apps cover all 21 categories of the store, including social networking, finance, business, and others. In each category, we picked up all the free apps when less than 100 of them are there, and top 100 otherwise.  Also from the iOS App Store, we collected 200 most popular apps, 40 each from ``All Categories'', ``Finance'', ``Business'', ``Social Networking'' and ``Productivity'', after removing duplications. The decrypted versions of these apps were extracted using Clutch~\cite{Clutch}.

All the apps were first quickly scanned to determine whether they utilize vulnerable services or channels, or export to their container directories. This was done by running the utility \texttt{otool} to extract Mach-O sections ``\texttt{\_\_objc\_selrefs}'' and ``\texttt{\_\_objc\_msgrefs}'' from each app's binary and look for the functions fingerprinting different services, channels and operations, as described in Section~\ref{subsec:detect}.  The apps discovered at this stage were further analyzed using Xavus for missing authentication operations. A problem is that Hopper does not support a batch mode.  To analyze an app, we had to manually load it into Hopper before it could be automatically evaluated by Xavus. This process was time-consuming, taking from about 1 to 30 minutes per app.  The developer of Hopper informed us that the batch mode will be supported in the near future.  For the time being, however, we could only analyze 200 randomly-chosen apps in the case that more were found to be associated with a channel or a service.

\vspace {3pt}\noindent\textbf{Vulnerable apps}. \ignore{Running Xavus on our app collection, we discovered a large number of vulnerable apps.  }Table~\ref{table:vulapps} summarizes our findings. Specifically, among all 1,612 MAC apps, 198 of them use the keychain.  Xavus did not find that any security check is performed by these apps between their claim of the keychain item and use of it to store sensitive data.  We further randomly chose 20 samples from the 198 apps and inspected them manually. It turns out that all of them can be easily attacked except todo Cloud and Contacts Sync For Google Gmail, which delete their current keychain items and create new ones before updating their data. Note that this practice (deleting an existing item) is actually \textit{discouraged} by Apple, which suggests to modify the item instead~\cite{kcitemdelete}.

\begin{table}[h]
\centering
\scriptsize
\begin{tabular}{|>{\centering}m{1.8cm}|>{\centering}m{1.3cm}|>{\centering}m{1.4cm}|c|}
\hline
\textbf{Channel} & \textbf{\begin{tabular}[c]{@{}c@{}}apps with the \\channel /total \end{tabular}} & \textbf{\begin{tabular}[c]{@{}c@{}}vulnerable/\\scanned\end{tabular}} & \textbf{\begin{tabular}[c]{@{}c@{}}exploitable/\\sampled\end{tabular}} \\ \hline
Keychain         & 198/1,612                                                            & 198/198                                                                     & 18/20                                                                        \\ \hline
NSConnection     & 58/1,612                                                              & 58/58                                                                       & 20/20                                                                        \\ \hline
Scheme (iOS)     & 138/200                                                              & 106/138                                                                     & 20/20                                                                        \\ \hline
Scheme(OS X)     & 982/1,612                                                            & 132/200                                                                     & 20/20                                                                        \\ \hline
BID              & 468/1,612                                                            & 468/468                                                                     & 20/20                                                                        \\ \hline
\end{tabular}
\caption{Vulnerable Apps}
\label{table:vulapps}
\end{table}

\begin{table*}[!t]
\centering
\begin{tabular}{|c|c|c|}
\hline
\textbf{XARA types}                                                              & \textbf{Secrets exposed}                                                                             & \textbf{Apps/Services affected}                                                                                                                  \\ \hline
\begin{tabular}[c]{@{}c@{}}Password Stealing\\ (keychain)\end{tabular} & \begin{tabular}[c]{@{}c@{}}passwords/\\ authentication tokens\end{tabular}                  & \begin{tabular}[c]{@{}c@{}}iCloud, Gmail, Google Drive, Facebook, Twitter,\\ any web account used in Chrome. \end{tabular} \\ \hline
IPC Interception                                                       & \begin{tabular}[c]{@{}c@{}}authentication tokens/\\ OS X username and password\end{tabular} & Keychain Access, 1Password, Evernote, Pushbullet.                                                                               \\ \hline
\begin{tabular}[c]{@{}c@{}}Scheme Hijacking \end{tabular}       & \begin{tabular}[c]{@{}c@{}}passwords/\\ authentication tokens\end{tabular}                  & \begin{tabular}[c]{@{}c@{}}Dropbox, Pinterest, Evernote, 1Password, \\Dashlane, Kindle, Instagram, Whatsapp. \end{tabular}  \\ \hline
\multirow{3}{*}{Container Cracking}                                    & email/cookies                                                                               & \begin{tabular}[c]{@{}c@{}}Foxmail,  App for Gmail, Mailtab for Gmail,\\ Mailtab for Outlook.\end{tabular}                      \\ \cline{2-3} 
                                                                       & \begin{tabular}[c]{@{}c@{}}notes/contacts/instant message pictures\end{tabular}          & Evernote, QQ, WeChat.                                                                                                           \\ \cline{2-3} 
                                                                       & cookies                                                                                     & \begin{tabular}[c]{@{}c@{}}Money Control, Inspire Finance Lite, Tumblr, AnyDo, Pocket. \end{tabular}                     \\ \hline
\end{tabular}
\caption{Examples of XARA Consequences}
\label{table:xaraconse}

\end{table*}

For the IPC on OS~X, we found that 58 apps use NSConnection\ignore{ and one runs XPC at the user mode}. All these apps were vulnerable\ignore{, except Apple Messages that operates XPC.  That app's BID starts with ``\texttt{com.apple}'', which is reserved for the Apple app and cannot be hijacked.  Therefore, its XPC server is not subject to the pre-emption attack (Section~\ref{subsec:IPC}).  However, the app can still share its object to an unauthorized client, due to the lack of protection on such an interaction.  From other detected apps,}. Again we sampled 20 and confirmed that all of them were indeed exploitable.

We did not find in our collection any free app using WebSocket.  However, there are popular paid apps claiming this channel. Particularly, 1Password is a leading paid app, which, as described in Section~\ref{subsec:IPC}, is completely vulnerable. Other examples include LastPass (a popular password management app), Adobe Creative Cloud (an Adobe service app) and LiveReload (for dynamic web content reloading).  These apps were all vulnerable to the attacks from malicious apps (Section~\ref{subsec:IPC}).

When it comes to Scheme, we discovered 982 MAC apps using this channel. From them, 200 apps were randomly picked and analyzed by Xavus, which reported that 132 were vulnerable. We further manually looked into 20 samples and successfully built end-to-end attacks.  The remaining 68 apps either use Apple reserved schemes or dynamically create their URLs from network traffic or other sources, which our current implementation cannot handle. Among the 200 iOS apps, 138 were detected to trigger URL schemes. 106 of them were reported to be vulnerable. Through random sampling, we confirmed that this finding is accurate. Also, those that could not be confirmed are very much in line with their MAC counterparts, either running reserved schemes or too complicated to analyze. Finally, 468 out of the 1,612 MAC apps were detected to write to their container directories, which can all be read by unauthorized apps hijacking their BIDs. Overall, at least 88.6\% of the scanned apps using these cross-app channels are vulnerable to XARA attacks.

\vspace {3pt}\noindent\textbf{Consequences}. Attacks on these vulnerable apps will have serious consequences. Table~\ref{table:xaraconse} lists some examples of the findings made in our research. Specifically, keychain credentials for high-profile services (e.g. iCloud, Gmail, Google Drive, Facebook, Twitter, LinkedIn, etc.) and any web accounts in Google Chrome are completely exposed. All their passwords and secret tokens can be collected by the adversary. Those vulnerable to the IPC interception include Keychain Access, Evernote, 1Password, Pushbullet, etc. Their sensitive data, such as authentication tokens and even current OS user's username and passwords are up for grabs.  The scheme vulnerability was found in 1Password, Dashlane, Evernote, Kindle, Adobe Revel, Wunderlist, etc., on OS~X, through which app users' credentials can be gathered. On iOS, popular apps like Pinterest, Instagram, U.S. Bank (banking), Citi Mobile (banking), PayPal, Amazon, WhatsApp, Dropbox, etc., were found to be exploitable. Their authentication tokens and other information can be stolen.



The BID confusion problem also has a significant impact. For example, our study shows that popular mail clients, such as App for Gmail, Mailtab for Gmail and Outlook, all expose MAC users' emails and their cookies to the app hijacking their BID.\ignore{ Also, Contacts Sync for Gmail was also detected to expose its cookie.}  Other apps that expose cookies include popular Finance apps Money Control and Inspire Finance Lite, as well as Tumblr, AnyDo, Pocket and more. Note that all the attack apps were successfully released by the Apple Stores. So, the security threats are indeed realistic.



\vspace {-5pt}
\subsection{Mitigation}
\label{subsec:defense}

Addressing the XARA problems is more difficult than it appears to be.\ignore{Changes may need to be made to the Apple access control framework to enforce the policies such as which NSConnection client is allowed to get the object from a specific server. Further complicating the situation is the fact that} Oftentimes, the OS itself does not know how to protect the resource of a third-party app.  Proper interfaces may need to be given to the app developers to let them specify and enforce their individual policies. A prominent example is the keychain, for which the OS is in no position to decide whether a set of attributes for retrieving an item should be used by one app but not others. Due to such complexity, these security weaknesses will likely be there for a while, before Apple figures out a way to work with the developers to fix them together. Indeed, since we reported the keychain issue to Apple in last October, so far, Apple did nothing except using a random \texttt{username} to patch some of its own apps, which turns out to be futile (Section~\ref{subsec:keychain}).

Given the challenges in finding a long-term solution\ignore{ for these problems}, it is important to have some protection in place to mitigate the threat\ignore{, informing Apple users whenever the security risks show up}. In this section, we describe a simple, lightweight scanner app, which automatically detects XARA attempts on OS~X.  As a third-party program running in the user land, this scanner can be easily deployed to provide the Apple user immediate protection.

\vspace {3pt}\noindent\textbf{Idea and implementation}. The idea of our XARA scanner is to inspect public information whenever a change to the system happens (e.g., write to a file, installation of a new app) to detect whether a service, resource or channel claimed by one app has been hijacked by another. This design enables the scanner to work efficiently and as we will show later, also effectively.  Specifically, our app registers file system event with API \texttt{FSEvents}, which is issued when a specific file has been modified. Specifically, the scanner monitors the keychain files under \texttt{/Library/Keychains/} and \texttt{$\sim$/Library/Keychains/}. Whenever they're modified, our app uses the API \texttt{SecItemCopyMatching} to find out whether a new item has been added, and if so further retrieves its ACL using \texttt{SecACLCopyContents} and inspects all the apps on the list. Typically, a system app does not share the list with a third-party app.  Once this is detected, the scanner notifies the user of the potential risk\ignore{ and suggests the protection measure that can be taken}. When it comes to a third-party app, all we can do is to build profiles for popular MAC apps through an offline analysis. Each profile contains the ACL an app is supposed to use, which is compared with the one retrieved from the keychain to detect an exploit attempt.


For Scheme and BID, our scanner keeps track of newly installed programs through the event API \texttt{FSEvents}. Whenever an app is installed, the scanner goes through its plist to find out whether the URL scheme it registers or the BIDs of its helper programs or XPC Services are in conflict with the ones already in the system.  Such a conflict indicates an exploit attempt, either from the new app or existing ones, and therefore triggers an alarm. Note that on scheme conflicts, even Apple does not know which app is legitimate to bond to a scheme. What Apple's OSes do is to arbitrarily associate the scheme with an app claiming that scheme (Section ~\ref{subsec:scheme}) while XARA scanner reports which app is associated with a scheme and the apps that fail to do so. 

For NSConnection, in the absence of the app developer's help, an exploit cannot be detected before it happens.  This is because whenever a third-party app claims an object name or requests an existing one, there is no reason to believe that the operation is illegitimate without the policies from the developer indicating that such resources should only be assigned to a specific app.  On the other hand, once the attack happens, it can be quickly detected. Our scanner redirects outputs of command \texttt{syslog -w} to get new system logs immediately after they are generated. Once it observes a failed attempt to register an existing object name or connect to an NSConnnection server that serves another app, an alarm is triggered, as such a conflict is not supposed to happen. This approach does not work on WebSocket, as the contest on a network port will not show up on the log.  The problem can be addressed using a heavyweight system-level solution, for example, running DTrace~\cite{DTRACE} to monitor the app's system calls.\ignore{ Such an approach, however, is very expensive and actually no more effective than our simple solution with regard to other security risks (keychain, Scheme, BID, NSConnection).  In our research, we implemented these user-land detection techniques into a native app.}

\vspace {3pt}\noindent\textbf{Evaluation}. We evaluated our implementation (a native app) against aforementioned XARA attacks (Section~\ref{sec:attack}) except that on WebSocket. Our scanner detected all the exploits on the keychain, URL schemes (on OS~X) and BIDs, before the malicious attempts could be executed.  For the NSConnection interceptions, it caught them from the events in the system log after the attacks happened. Note that since such contention of app-specific resources, channels or services does not exist during the system's normal operations, our scanner will not cause a false alarm, though it might miss some exploit attempts. We further measured its performance on a MacBook Pro (Mid 2014 model, 2.6 GHz Intel i5, 8 GB memory, SSD), under OS~X 10.10.2. It utilizes no more than 0.2\% of CPU during operations. \ignore{More details are available in Section~\ref{appendix:overhead} (see Appendix)}\ignore{,  The memory cost during operations remains almost constant at around 50 MB even with over 1,600 apps installed on the Mac. The performance overhead is also negligible, below 0.97\% during operations (Section~\ref{appendix:overhead})}.

\vspace {-8pt}
\section{Lessons Learnt}
\label{sec:discuss}

Almost all the XARA weaknesses we discovered in this research come from Apple's unique design of cross-app resource sharing and communication mechanisms, e.g., keychain for sharing passwords, BID based separation, NSConnection for distributing objects and URL scheme for app invocation (different from Android). Other XARA problems, i.e., the WebSocket issues may also exist on other OSes, such as Windows and Android. This demonstrates that the XARA weakness is indeed pervasive and serious\ignore{, across different platforms}.  A natural question here is how those problems have been introduced and what we can learn from them.  In the section, we try to answer the question, presenting the insights gained from analyzing those vulnerabilities and the principles for designing a securer system.

\vspace {3pt}\noindent\textbf{Insights}. The fundamental cause for the XARA flaws is unprotected cross-app resource sharing and communication. Comparing OS~X with iOS, the latter is relatively securer simply because it does not support credential sharing (among different apps) through a keychain item and sub-target sharing (e.g., framework) through containers, nor does it provide any complicated IPC mechanism like NSConnection. For every avenue opened across apps, proper authentication should always be in place. Otherwise, an XARA risk may show up.

Apparently, XARA is an instance of the classic Unverified Ownership or Resource Squatting problem~\cite{cwe283, cwe377, vsj_2012_usenixsec}, in which software fails to verify which party owns a piece of critical resource. The unique challenge in addressing the issue, however, is that when it comes to the interactions across third-party apps, less clear are who should perform the verification and how to do so. For example, when an app deposits the user's credential to another party's keychain item on OS~X, as long as it is indeed on the item's ACL, the operating system is not at a position to judge the legitimacy of the operation, since this is allowed for credential sharing. As another example, neither OS~X nor iOS has any idea which app is entitled to a specific URL scheme. The authentication here (on the owner of the keychain item or the scheme before delivering data to it) can only be performed by the app. Yet, the OS provider still has the responsibility to assist the app developer in implementing such protection (e.g. providing proper APIs) and further verify its presence in her app, which is essential for fostering a secure ecosystem.

Following we summarize the above insights into three key principles for avoiding XARA hazards in cross-app interactions.

\vspace {3pt}\noindent\textbf{Design Principle 1:} \textit{Determine what the OS can protect and what it cannot for every cross-app channel}. When a new way for cross-app resource sharing or communication is provided, the OS designer always needs to determine whether authenticating the parties involved can be done at the OS level or only by individual apps. \textit{The OS needs to address the security issues whenever possible to ensure the effectiveness of the protection and makes it clear what should be taken care of by the app developer.} Among all the XARA cases we discovered, container sharing should be fully secured by the OS: it has sufficient information to decide what is allowed to share (e.g., framework) and what is not (e.g., the helper program's directory). On the other hand, the keychain item, parties in an IPC and the scheme owner often can only be checked by the app. It is important to identify how to divide the responsibilities of security protection at the early stage of developing cross-app channels.

\vspace {3pt}\noindent\textbf{Design Principle 2:} \textit{Inform the app developer required app-side security checks and provide means to do so}.  Whenever a cross-app channel is found to need the app developer's involvement to secure, the OS provider should explicitly inform the developer what she is supposed to do and provide proper technical supports. Our study shows that this is exactly what Apple falls short. Oftentimes, it does not offer any APIs for the required authentication: examples include NSConnection, Keychain and WebSocket, etc.  Even when the API is available, e.g., for finding the app to be launched through a scheme, rarely have we found any instructions for the developer to do that. In the absence of such supports, XARA flaws become inevitable.

\vspace {3pt}\noindent\textbf{Design Principle 3:} \textit{Detect missing security checks at the app store}. Even with the proper information and technical means,  we believe that the OS provider can do more to help the app developer and secure its app ecosystem. What can be leveraged here is how the apps are disseminated today: they are mainly downloaded and installed from a centralized app store under the control of the provider, which enables the provider to make vulnerability detection part of its app vetting process. This is complete feasible, given the fact that today the Apple Store takes more than a week to approve an app while the automatic tools like Xavus can be built to detect missing authentication within the app in minutes.



\vspace {-5pt}
\section{Related Work}
\label{sec:relatedwork}
\vspace {-2pt}

\vspace {3pt}\noindent\textbf{XARA attacks on Android}. Security flaws related to XARA have been discovered on Android, e.g., different types of confused deputy problems within Android apps~\cite{felt2011permission, Chin:2011:AIC:1999995.2000018, grace2012systematic, Lu_chex:statically}. Most relevant to our work is the prior research on mobile origin crossing~\cite{wang2013unauthorized}, which reports an attack that a malicious Android app registers the scheme of a URL not meant for invoking apps and runs it in a browser to get a Facebook token. This problem is \textit{not} a scheme hijacking, since the scheme here is not associated with any legitimate app. Actually, preempting another app's scheme is hard on Android because whenever there are two apps registering the same scheme, Android always notifies the user and let her make the decision. Also such a problem has already been fixed by Facebook, yet the scheme hijacking is an issue they are not aware of. Also related is the study on the Pileup~\cite{Xing:2014:UYA:2650286.2650760} attacks, in which a malicious app can gain an elevated privilege through a system upgrade. The problem here is not in the design of isolation protection but the mechanism to grant an app additional permissions, which has been circumvented in the upgrade process.

\vspace {3pt}\noindent\textbf{Security on the Apple platforms}. Compared with Android, the Apple platforms are much less studied in terms of their security protection. A technical blog~\cite{scheme2010} talked about insecure handling of schemes in the invoked apps on iOS, which is not the scheme hijacking on both OS~X and iOS, as discussed in this paper. Prior academic research almost solely focuses on various techniques to bypass the security checks on iOS private APIs~\cite{Wang:2013:JIB:2534766.2534814, DBLPHanKYBDGLZ13} and use of them to propagate malware infections~\cite{tieleiwang}.  Understanding the security implications of Apple's inter-app interactions and sandbox design has never been done before. Simultaneously and independently, Fireye found the risk of hijacking iOS schemes and put a blog online~\cite{fireeye}. However, they just briefly discuss this security risk without giving much detail, with a demo that apparently shows a simple phishing attack. By comparison, our work is much more thorough, deeper and broader. We built end-to-end attacks on several high-impact apps (e.g., Facebook, Pinterest, etc.), identified the impacts of the threat over a thousand apps, and more importantly demonstrate that the attacks can be made stealthy (through different man-in-the-middle tricks on MAC OS and iOS, passing the stolen token to the victim app, to completely conceal the attack), which is nontrivial (see Section~\ref{subsec:scheme}). Also we completely circumvented the restrictive security checks of the Apple Stores: actually, our attack apps were approved by the App Store on January 23, 2015, almost one month earlier than the blog (February 19), which did not mention any study on the protection provided by the App Store.  Further, we discovered that the problem exists on \textit{both iOS and OS~X} and different strategies these OSes took to resolve conflicts in Scheme claims (Section~\ref{subsec:scheme}), which is important to the success of the attack. \ignore{Note that our Youtube demo for the attack on OS~X schemes~\cite{supporting} was uploaded on February 5 before the blog, which \textit{only} mentions the attack on iOS schemes. }Finally, we developed techniques for automatically detecting such exploits and mitigating this risk.

Related to Xavus is PiOS~\cite{egele11:pios}, a general-purpose code analysis tool for iOS apps, which has not been made publicly available so far.  By comparison, our approach was designed specifically for detecting XARA flaws within both MAC OS and iOS apps.

\section{Conclusion}
\label{sec:conclude}
In this paper, we identify a new category of security weaknesses, called XARA, that pose a serious threat to the app isolation protection on modern OSes. Our study on the threat over the Apple platforms, the first of this kind, reveals its pervasiveness and significant impacts: critical system services and channels, including the keychain, WebSocket, NSConnection and Scheme, can all be exploited to gain access to other apps' resources, and even the Apple Sandbox on OS~X can be cracked, exposing an app's container directory to the unauthorized party. The consequences of these attacks are serious, including leaks of user passwords, secret tokens and all kinds of sensitive documents. Our research shows that fundamentally the problem comes from lack of authentication during app-to-app and app-to-system interactions, and further proposes new techniques to detect and mitigate such a threat. This preliminary effort contributes to a better understanding of this understudied security problem, an important step for building a more effective app isolation mechanism on future OSes.


\begin{thebibliography}{100}

\bibitem{1pass}
1Password - Password Manager and Secure Wallet. 
https://itunes.apple.com/us/app/
1password-password-manager/id443987910?mt=12.

\bibitem{Clutch}
Clutch. https://github.com/KJCracks/Clutch.

\bibitem{CocoaHTTPServer}
CocoaHTTPServer. https:
//github.com/robbiehanson/CocoaHTTPServer.

\bibitem{machnotimp}
Code Signing Services Reference. \url{https://developer.
apple.com/library/mac/documentation/Security/Reference/CodeSigningRef/index.html#//apple_
ref/doc/constant_group/Attribute_Selector_
Dictionary_Keys}.

\bibitem{cwe283}
CWE-283: Unveried Ownership. https:
//cwe.mitre.org/data/definitions/283.html.


\bibitem{cwe377}
CWE-377: Insecure Temporary File. https:
//cwe.mitre.org/data/definitions/377.html.

\bibitem{DTRACE}
DTRACE. https://developer.apple.com/library/
mac/documentation/Darwin/Reference/ManPages/
man1/dtrace.1.html.

\bibitem{hopper}
Hopper V3, the OSX and Linux Disassembler.
http://www.hopperapp.com/.

\bibitem{scheme2010}
Insecure Handling of URL Schemes in Apple's iOS.
http:
//software-security.sans.org/blog/2010/11/08/
insecure-handling-url-schemes-apples-ios/.

\bibitem{fireeye}
 iOS Masque Attack Revived: Bypassing Prompt for
Trust and App URL Scheme Hijacking.
\url{https://www.fireeye.com/blog/threat-research/
2015/02/ios_masque_attackre.html}.

\bibitem{kcitemdelete}
 Keychain Services Reference. \url{https://developer.
apple.com/library/mac/documentation/Security/
Reference/keychainservices/index.html#//apple_
ref/c/func/SecKeychainItemDelete}.

\bibitem{MachO}
OS X ABI Mach-O File Format Reference. \url{https://
developer.apple.com/library/mac/documentation/
DeveloperTools/Conceptual/MachORuntime/index.
html#//apple_ref/doc/uid/TP40000895}.

\bibitem{templatecode}
OS X Keychain Services Tasks. \url{https://developer.
apple.com/library/ios/documentation/Security/
Conceptual/keychainServConcepts/03tasks/tasks.
html#//apple_ref/doc/uid/TP30000897-CH205-TP9}.

\bibitem{QtWebKit}
QtWebKit.
http://qt-project.org/wiki/QtWebSockets.

\bibitem{supporting}
Supporting materials.
https://sites.google.com/site/xaraflaws/.

\bibitem{WebSocket}
The WebSocket API.
http://www.w3.org/TR/websockets/.

\bibitem{Chin:2011:AIC:1999995.2000018}
E. Chin, A. P. Felt, K. Greenwood, and D. Wagner.
Analyzing inter-application communication in android.
In Proceedings of the 9th International Conference on
Mobile Systems, Applications, and Services, MobiSys
'11, pages 239-252, New York, NY, USA, 2011. ACM.

\bibitem{Davi:2010:PEA:1949317.1949356}
L. Davi, A. Dmitrienko, A.-R. Sadeghi, and
M. Winandy. Privilege escalation attacks on android.
In Proceedings of the 13th International Conference on
Information Security, ISC'10, Berlin, Heidelberg,
2011. Springer-Verlag.

\bibitem{egele11:pios}
M. Egele, C. Kruegel, E. Kirda, and G. Vigna. PiOS:
Detecting Privacy Leaks in iOS Applications. In
Proceedings of the Network and Distributed System
Security Symposium (NDSS), San Diego, CA, February 2011.

\bibitem{felt2011permission}
A. P. Felt, H. J. Wang, A. Moshchuk, S. Hanna, and
E. Chin. Permission re-delegation: Attacks and
defenses. In USENIX Security Symposium, 2011.

\bibitem{grace2012systematic}
M. Grace, Y. Zhou, Z. Wang, and X. Jiang.
Systematic detection of capability leaks in stock
android smartphones. In the 19th Annual Symposium
on Network and Distributed System Security, 2012.

\bibitem{DBLPHanKYBDGLZ13}
J. Han, S. M. Kywe, Q. Yan, F. Bao, R. H. Deng,
D. Gao, Y. Li, and J. Zhou. Launching generic attacks
on ios with approved third-party applications. In
ACNS, 2013.

\bibitem{vsj_2012_usenixsec}
Hayawardh Vijayakumar and Joshua Schiman and
Trent Jaeger. STING: Finding Name Resolution
Vulnerabilities in Programs. In Proceedings of the 21st
USENIX Security Symposium (USENIX Security
2012), August 2012.

\bibitem{Lu_chex:statically}
L. Lu, Z. Li, Z. Wu, W. Lee, and G. Jiang. Chex:
statically vetting android apps for component
hijacking vulnerabilities. In In Proc. of the 2012 ACM
conference on Computer and communications security,
CCS 2012, ACM.

\bibitem{wang2013unauthorized}
R. Wang, L. Xing, X. Wang, and S. Chen.
Unauthorized origin crossing on mobile platforms:
Threats and mitigation. In the 20th ACM conference
on Computer and communications security. ACM,
2013.

\bibitem{tieleiwang}
T. Wang, Y. Jang, Y. Chen, S. Chung, B. Lau, and
W. Lee. On the feasibility of large-scale infections of
ios devices. In 23rd USENIX Security Symposium
(USENIX Security 14), San Diego, CA, Aug. 2014.
USENIX Association.

\bibitem{Wang:2013:JIB:2534766.2534814}
T. Wang, K. Lu, L. Lu, S. Chung, and W. Lee. Jekyll
on ios: When benign apps become evil. In Proceedings
of the 22Nd USENIX Conference on Security, SEC'13,
pages 559-572, Berkeley, CA, USA, 2013. USENIX
Association.

\bibitem{Xing:2014:UYA:2650286.2650760}
L. Xing, X. Pan, R. Wang, K. Yuan, and X. Wang.
Upgrading your android, elevating my malware:
Privilege escalation through mobile os updating. In
Proceedings of the 2014 IEEE Symposium on Security
and Privacy, 2014.

\end{thebibliography}
\end{document}